\documentstyle[aps,pre,multicol,epsfig]{revtex}
\draft
\input epsf
\begin{document}
\title{\bf Stretched exponential relaxation in a
diffusive lattice model.}
\author{C. Fusco${}^\dagger$, 
P. Gallo${}^\dagger$, A. Petri${}^\ddagger$ and M. Rovere${}^\dagger$
}
\address{${}^\dagger$ Dipartimento di Fisica, Universit{\`a} Roma
Tre, \\ 
and Istituto Nazionale per la Fisica della Materia,  Unit{\`a} di Ricerca
Roma Tre, \\ Via della Vasca Navale 84, I-00146 Roma, Italy \\
${}^\ddagger$ Consiglio Nazionale delle Ricerche,
Istituto di Acustica ``O. M. Corbino'',\\
Area della Ricerca di Roma Tor Vergata,\\
Via del Fosso del Cavaliere 100, 00133 Roma, Italy
}

\maketitle

\begin{abstract} 

We studied the single dimer dynamics in a lattice
diffusive model as a function of particle density in the
high densification regime. The mean square displacement 
is found to be subdiffusive both in one and two dimensions.
The spatial dependence of
the self part of the van Hove correlation function displays as function
of $r$ a single peak and signals a dramatic slow down
of the system for high density. The self intermediate scattering 
function is fitted to the Kohlrausch-Williams-Watts law. The exponent
$\beta$ extracted from the fits is density independent while the relaxation
time $\tau$ follows a scaling law with an exponent $2.5$.

\end{abstract}
\begin{multicols}{2}
\section{Introduction}

Lattice-gas models of dense fluids have been 
explored ~\cite{b.Kob2,b.Kob3,b.coniglio}
in order to understand the microscopic details of their 
slow dynamics and to compare them with the behavior of 
other disordered systems like structural glass
formers, spin glasses and colloids. 

Among the lattice models {\em Random Sequential Adsorption} 
(RSA) of particles on a substrate is of particular interest. 
RSA has been invented to mimic many experimental situations
like chemisorption on crystal surfaces, reactions on polymer
chains, adsorption of macromolecules and colloidal particles, monolayer and
multilayer growth processes (for an
an extensive review see~\cite{b.Evans,b.Privman1}). 
In particular RSA with 
added Diffusion (RSAD) has been considered a model which 
captures the main features of the compaction dynamics in granular 
matter~\cite{b.deOliveira}. 
The description of the relaxation processes in the RSAD model 
has been investigated through the behavior of the
time density correlation function in a limited number of studies and only
during the deposition process \cite{stint1,stint2}. 

In this paper we explore by computer simulation the equilibrium 
dynamical properties of a lattice gas of dimers as a function of density
in two dimensions.
The starting configurations have been produced with a RSAD model where 
by deposition and diffusion of particles we reach the desired value for 
the density. 
In this way, due to the geometrical frustration, as the dimer density 
increases trapping configurations that only 
depend on the local environment \cite{b.Fusco} are created in the system.
The dynamical properties are studied by switching the deposition off. 

In the next section we describe the model and the details of our
Monte Carlo simulation. In the third section we present our results
for the single dimer dynamics as a function of 
particle density. In particular we analyze the mean square displacement, 
the Van Hove self correlation function and its space Fourier
transform, the self intermediate scattering function. 
Last section is devoted to concluding remarks. 

\section{Model and Computational Details}

In order to investigate the dynamical properties of dimers on a square
lattice with fixed density we have generated the starting configurations
through a RSA model of dimers with diffusion 
in two dimensions (RSAD$2d$)~\cite{b.Fusco,b.Grigera} 
by Monte Carlo (MC) simulations.
We performed also simulations of the same model in one dimension (RSAD$1d$)
for comparison.
We have considered a 
square lattice of $N=L\times L$ sites, with periodic boundary 
conditions in both directions. 
Two kinds of dynamical processes can occur in the system, 
namely deposition of a dimer on an empty pair of sites and diffusion 
of a deposited dimer on an empty first neighbor site; no overlap
of dimers is allowed. The lattice is initially  empty and at 
each MC step one 
deposition event (or one diffusion event) is selected with probability $p$
(or $1-p$), where $0<p\le 1$. 
Dimers can be adsorbed on the lattice in two 
possible orientations (horizontal and vertical) and  can diffuse along 
their own axis. 
In the case of deposition, the dimer can occupy a
randomly selected pair of sites if they are both empty, 
otherwise the attempt is rejected. In the case of diffusion the 
randomly selected horizontal (vertical) particle attempts to move left or 
right (up or down), with equal probability. 
A diffusion event to the right (left) or up (down) is possible only if the 
chosen pair of sites is occupied and the corresponding right (left) 
or up (down) neighbor site is empty, otherwise the attempt is rejected.
As usual the MC time unit corresponds to $N$ deposition/diffusion attempts.
We checked that 
finite-size effects are negligible for $L\ge 40$, so in the following we
analyze a square lattice with $L=40$. 
We produced 11 starting configurations at densities
$\rho=0.60,0.70,0.75,0.80,0.88,0.94,0.96,0.98,0.99,0.995,0.998$.
For each configuration production runs at constant
density have been obtained by switching the deposition
events off. In this way we are able to follow 
the dynamics of the system in an equilibrium situation.
In fact, the equilibrium distribution for this system, 
where all allowed configurations have the same weight,
is a constant. Therefore the detailed balance condition
reduces to two requirements: transitions
to forbidden states are not allowed, and transition from an allowed
state to another has the same probability as the reverse one. 
Both these requirements are satisfied by the adopted algorithm.

A more complete investigation of the irreversible dynamics of these models
and discussion of the role of the dimensionality is 
presented elsewhere~\cite{b.Fusco}.

\section{Single Dimer Dynamics}

In this section we investigate the dynamical behavior of this lattice  
model through the single dimer density correlators.
We have chosen relatively large values of the density 
($\rho\ge 0.60$), since we are
interested in the diffusive dynamics of the system in the high densification
regime where slow dynamics is more likely to occur. 
The time over which the diffusive motion 
of particles has been followed reaches in most cases $10^{6}$ MC time units
(because of the  statistical noise in the last decade 
only $t$ up to $10^{5}$ is shown in the plots). 

\subsection{Diffusion properties}

\begin{figure}
\centering\epsfig{file=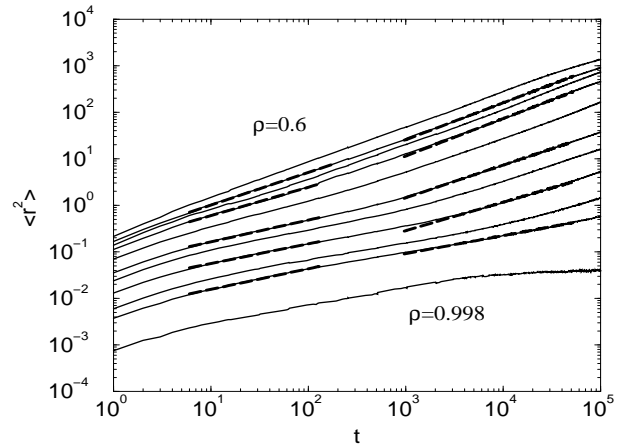, width=8cm, height=6cm}
\caption{Log-log plot of the temporal evolution 
of the MSD  for several different densities:
$\rho=0.60,0.70,0.75,0.80,0.88,0.94,0.96,0.98,0.99,0.995,0.998$
from top to bottom. Long dashed lines are the fit to power law  behaviour
in the two time regions from $5$ to $120$ MC timesteps and from 
$1000$ to $60000$ MC timesteps. For the sake of clarity fits are reported
only for even curves. The exponents extracted from
the fits are reported in table \protect\ref{table1}.}
\protect\label{f.1}
\end{figure}

We analyze in the following the mean square displacement (MSD)
of the particle, $<r^{2}(t)>$, defined as 
\begin{equation}
\label{e.r2}
<r^{2}(t)>=\frac{1}{N_{part}}<\sum_{i=1}^{N_{part}}|{\mathbf{r}}_{i}(t)-
{\mathbf{r}}_{i}(0)|^{2}>,
\end{equation}
where $N_{part}=N \cdot \rho /2 $ is the 
number density, $\mathbf{r}_{i}$ is 
the position of particle $i$ at time $t$ and $<>$ denotes a time average.
In fig.~\ref{f.1} we show a log-log plot 
$<r^{2}(t)>$ for RSAD$2d$ for the different 
densities investigated.
After an initial transient, which is more pronounced  
at high densities, the
temporal behavior of the MSD can be fitted with two power laws.
In the first time region  
$<r^{2}(t)>\sim t^{\delta_{1}}$. 
In the intermediate time region for the lower densities the MSD
shows a bent followed by a slight increase in the exponent, i.e. 
$<r^{2}(t)>\sim t^{\delta_{2}}$ with $\delta_{2}>\delta_{1}$, 
where the exponents depend upon density. 
The bent becomes more evident as the density increases.
Values of  
$\delta_{1}$ and $\delta_{2}$ are reported in table \ref{table1}.
\begin{table}
\centering
\caption{Values of the exponents extracted from the fit to
the power law of the MSD plotted in Fig.1.}
\begin{tabular}{|c|c|c|c|}
Density & $\delta_1$&$\delta_2$&$\delta_2/\delta_1$\\ 
\hline 
0.60 &           0.75 &           0.80 &         1.07 \\
0.70  &          0.72 &           0.80 &         1.11 \\
0.75   &         0.67 &           0.79 &         1.18 \\
0.80   &         0.64 &           0.79 &         1.23 \\
0.88   &         0.60 &           0.74 &         1.23\\
0.94   &         0.54 &           0.70 &         1.30\\
0.96   &         0.52 &           0.67 &         1.29\\
0.98   &         0.51 &           0.60 &         1.18\\
0.99   &         0.51 &           0.48 &         0.94\\
0.995  &         0.51 &           0.38 &         0.75\\
\end{tabular}
\label{table1}
\end{table}
They are less than $1$ for 
all the densities considered: in particular $\delta_{1}$ and $\delta_{2}$ do 
not exceed respectively $0.75$ and $0.80$,
which are the values found for the lowest density, i.e. $\rho=0.60$. 
The difference between the two exponents is more 
marked for intermediate densities ($0.80<\rho<0.96$). For the 
highest density ($\rho=0.998$) the MSD shows a plateau for large times.
In the time region investigated none of the curves shown attains
the Brownian diffusive regime, $\delta_2=1$.
 
These findings can be interpreted in terms of the density characterizing the 
degree of motion of particles like the thermal energy in thermal systems. 
For low densities single particles can move freely and the diffusive dynamics is
almost independent of time. Increasing the density
particles remain trapped in local transient 
configurations~\cite{b.Fusco,b.Grigera}
and diffusion slows down. 
For very high density the trapping configurations are frozen and 
the system does not evolve any more on the time 
scale of our simulations. The existence of trapping configurations and
subdiffusive behavior is in general linked to the presence of 
disorder~\cite{b.Haus} that in  the present case can be ascribed to 
the random barriers 
associated with the local environment determined by the geometrical 
constraints.
The particle has to overcome these barriers and spend a long time before
detrapping (see~\cite{b.Avramov} for an analogous case in a random walk 
model). 

These dynamical features are not 
present in the RSAD$1d$ model, where 
we have found a MSD with a density 
independent slope for this case. Nevertheless the 
behavior is subdiffusive, with an exponent
$\delta=0.5$ constant over the whole time window investigated. 
        
\subsection{Density correlator}

For a closer analysis of the motion of the dimer we have considered the 
self part of the van Hove correlation function 
(SVHCF) $G_{s}(r,t)$,  defined as
\begin{equation}
\label{e.vanhove}
G_{s}(r,t)=\frac{1}{N_{part}}<\sum_{i=1}^{N_{part}}
\delta(r-|{\mathbf{r}}_{i}(t)-{\mathbf{r}}_{i}(0)|)>,
\end{equation}  
which gives the probability of finding a particle around a distance
$r$ at time $t$, given that the same particle 
was at the origin at time $t=0$. 
The MSD analyzed in the previous subsection
is the second spatial moment of the SVHCF. 
Nonetheless, in disordered diffusive
systems that do not have a simple behavior, 
the SVHCF might contain additional information on the microscopic
relaxation mechanisms of the system, that could be smeared out in the MSD.

\begin{figure}[h]
\centering\epsfig{file=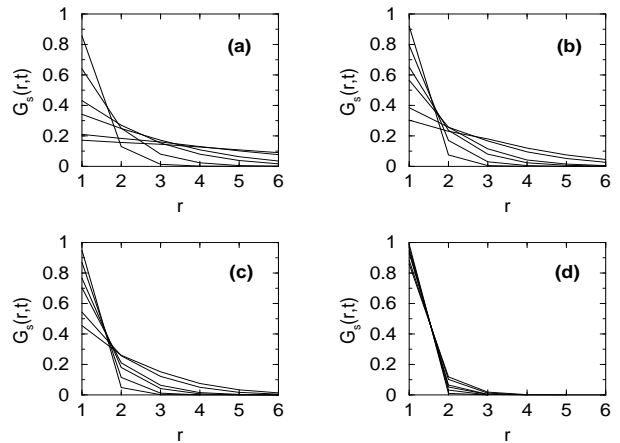, width=8cm, height=6cm}
\caption{Spatial dependence of the SVHCF
$G_{s}(r,t)$ for different densities: 
$\rho=0.60$ (a), $\rho=0.80$ (b), $\rho=0.88$ (c) and $\rho=0.98$ (d). 
The curves in each graph correspond to various times: 
$t=3,15,63,127,511,1023$ (in MC time units) from top to bottom on the left 
side of each graph. Distances on the horizontal axis are measured in 
lattice spacing units.}
\protect\label{f.2}
\end{figure}

The SVHCF is shown for our system 
in fig.~\ref{f.2} for four densities at different times. 
We have calculated the one dimensional SVHCF
along the horizontal or the vertical direction.
In fig.~\ref{f.2} we show the SVHCF averaged over the two directions.
The self correlation functions shows a single peak 
centered around zero for all the times
investigated. The appearance of more than one peak 
at a given time would have evidenced
the existence of more than one, well distinct, relaxation mechanism
for that time. 
The single peaked structure confirms the dynamics
to be mastered only by a single mechanism related to the existence of
high entropic barriers generated by configurations in which a 
particle is locally trapped. 
It is also seen that at low density the SVHCF is 
localized at short distances
(typically $1\le r\le 2$) for short times and tends to delocalize for large 
times, since it is able to explore more configurations. 
At  higher densities we observe a ``clustering''
of the various curves for $1\le r\le 2.5$ (this effect is particularly evident
in fig.~\ref{f.2}(d)): this is the signature of the fact that the movement of 
particles drastically slows down at such densities. 

\subsection{Intermediate scattering function}

\begin{figure}[h]
\centering\epsfig{file=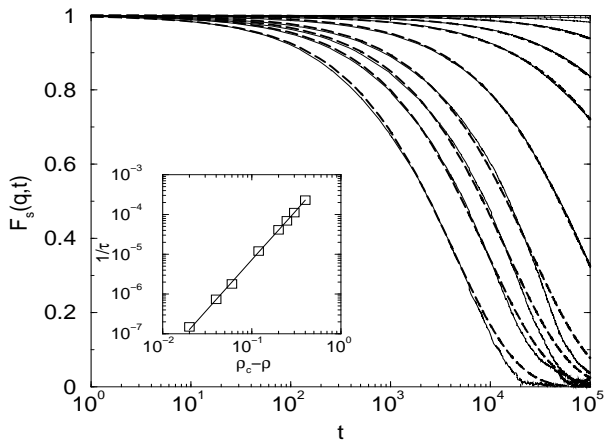, width=8cm, height=6cm}
\caption{Temporal evolution of the SISF $F_{s}(\mathbf{q},t)$ for
${\mathbf{q}}=(2\pi(L-1)/L,0)$. 
The solid curves correspond to the result of the
simulations for different densities: 
$\rho=0,60,0.70,0.75,0.80,0.88,0.94,0.96,0.98,0.99,0.995,0.998$.
The long-dashed curves are fits to the data according to 
eq.~(\ref{e.corfit}).
In the inset it is shown the inverse relaxation time vs. density: 
the squares are
the parameters obtained from the fit to eq.~(\ref{e.corfit}) 
for various density,
while the solid curve is the best fit according to eq.~(\ref{e.taufit}).}
\protect\label{f.3}
\end{figure}

The space Fourier transform of the SVHCF, i.e. the {\em Self Intermediate 
Scattering Function} (SISF) $F_{s}({\mathbf{q}},t)$, is given by
\begin{equation}
\label{e.cor}
F_{s}({\mathbf{q}},t)=\frac{1}{N_{part}}\sum_{i=1}^{N_{part}}
\exp \left[ i{\mathbf{q}}\cdot ({\mathbf{r}}_{i}(t)-{\mathbf{r}}_{i}(0)) 
\right],
\end{equation}
where ${\mathbf{q}}=(q_{x},q_{y})$ is the discrete wave vector 
($q_{x,y}=2\pi n_{x,y}/L$ with $n_{x,y}$ an integer between $0$ and $L-1$). 
For all the densities considered in the simulations $F_{s}(q,t)$ is found to
decay more slowly the smaller $q$ is.
We show in fig.~\ref{f.3} data for 
${\mathbf{q}}=(2\pi(L-1)/L,0)$ at several different densities. 
This $q$ value is chosen of the order of the inverse of the
typical size of the trapping region, which we estimate to be of the order
of $2-3$ lattice spacing~\cite{b.Fusco}. 
As it can be seen, the entire dynamical behavior at 
low-intermediate densities is observed
in the time window of our simulations, 
since the SISF decays to zero for such densities. 
For high values of the density
only a part of the SISF can be observed, due to the very slow
dynamical relaxation and the freezing of particles which is also evident in
the MSD. The  short-intermediate time behavior  of SISF can be fitted with a 
Kohlrausch-Williams-Watts law (KWW)
\begin{equation}
\label{e.corfit}
F_{s}(q,t)\simeq \exp[-(t/\tau)^\beta], \qquad 0<\beta\le 1.
\end{equation}
The curve resulting from this fitting, also shown  in fig.~\ref{f.3},  
reproduces the data
obtained from MC simulations quite accurately for several decades.
A deviation of MC results from eq.~(\ref{e.corfit}) is observable only
in the tails of the correlators. 
The collapse of 
correlators onto a single curve when plotted versus $t/\tau(\rho)$
demonstrates  the existence of a well 
defined scaling law on several decades for this kind of systems. 
In the range of density considered the stretching exponent $\beta$ is 
essentially constant, with a value $\sim 0.68$, the
density dependence being reflected only in the relaxation time $\tau$. 

Furthermore we have found that the behavior of $\tau$ versus $\rho$ follows 
the scaling law
\begin{equation}
\label{e.taufit}
\tau^{-1}\simeq A(\rho_{c}-\rho)^\gamma.
\end{equation} 
This fit is shown in the inset of fig.~\ref{f.3} together 
with the points determined from 
the parameters of eq.~(\ref{e.corfit}). 
As it can be seen from the figure, 
eq.~(\ref{e.taufit}) applies satisfactorily in the entire range of densities 
considered. Our estimates yield  $A\simeq 0.0023$, $\gamma\simeq 2.5$ and 
$\rho_{c}\simeq 1.0$. Given the uncertainties in the fit, 
this value found for $\rho_{c}$ cannot
be distinguished from the maximal density reachable in this system,
which has been found  to be $ \approx 0.998$  in numerical simulations 
(see~\cite{b.Fusco,b.Grigera}).
This would suggest that there is no dynamical transition in this model, i.e.
no critical density (lower than the maximal density) exists at which a 
structural arrest of the system occurs.
It is important to stress that the relaxation time obeys the power law 
eq.~(\ref{e.taufit}), which has the same kind of 
scaling found in many glassy systems and is the signature of a sluggish 
dynamical behavior. 
The existence of a power law 
behavior for $\tau$ as a function of density and the validity of 
KWW law are novel and interesting points 
for systems of this kind.  
It would be interesting to investigate in the future
the $\tau$ and $\beta$ dependence on the wavevector to check 
on the range of validity of both the KWW law and of the power law. 

\section{Conclusions}

We presented a Monte Carlo study of the single dimer density correlators for
a randomly diffusive model of dimers on a square lattice
as function of density. For each density the starting configurations
have been generated through a RSAD2d model.
We found that 
the dynamical behavior is related with the system dimensionality and 
that the diffusive dynamics is non Brownian for the time regimes investigated.
In spite of its simplicity the model considered shows a quite complex
dynamical behavior, and in particular some features resembling 
that of the slow relaxation of structural glass
formers are found. 

The most interesting results came from the SISF analysis. We have 
found that this correlator in the $q,t$ space has a well defined 
scaling behavior with density that can be described
with a high degree of accuracy by the KWW function in most
of the time region investigated. This function
is well known to reproduce many features of disordered systems, 
and is connected to fractality in configurational 
space~\cite{b.Gotze,b.Jund}.
The exponent of the stretched exponential function $\beta$ is
density independent and the relaxation time follows a power law
with an exponent $\gamma$ in the range of those of 
structural glass formers \cite{b.Gotze}.
The power law predicts a divergence of the relaxation time for
density around one. 
From these findings it can be consistently derived that for this non thermal 
system the role of thermal energy  is played by the density.      

We correspondingly find a subdiffusive behavior in the MSD. 
The slow down  of the system that becomes more significant as 
density is increased can be attributed to  transient 
trapping configurations. 
We point out  that  with respect to 
lattice-gas models  of the kind considered in ref.~\cite{b.Kob3} 
we did not introduce any explicit constraint to induce the trapping of the
dimers. In our system in fact it naturally arises 
because of geometrical constraints.  

\acknowledgments

We thank V. Loreto and A. Barrat for useful discussions.

\end{multicols}

\end{document}